\documentclass{easychair}

\usepackage{titlesec}
\titlespacing{\paragraph}{0pt}{\medskipamount}{*1.5}

\usepackage{mdwlist} 
\usepackage{mathtools} 

\usepackage{agda}
\AgdaNoSpaceAroundCode{}
\newcommand{\AF}[1]{\AgdaFunction{#1}}
\newcommand{\AR}[1]{\AgdaField{#1}}

\usepackage{newunicodechar}
\newunicodechar{→}{\ensuremath{\to}}
\newunicodechar{⇒}{\ensuremath{\Rightarrow}}
\newunicodechar{λ}{\ensuremath{\lambda}}
\newunicodechar{ƛ}{\ensuremath{\lambda}}
\newunicodechar{∀}{\ensuremath{\forall}}
\newunicodechar{∃}{\ensuremath{\exists}}
\usepackage{graphics}
\newcommand\new{\reflectbox{\ensuremath{\mathsf{N}}}}
\newunicodechar{И}{\new}
\newunicodechar{≡}{\ensuremath{\equiv}}
\newunicodechar{≈}{\ensuremath{\approx}}
\newunicodechar{≐}{\ensuremath{\doteq}}
\newunicodechar{∷}{::}
\newunicodechar{◇}{+\!\!\!+}
\newunicodechar{∈}{\ensuremath{\in}}
\newunicodechar{∉}{\ensuremath{\notin}}
\newunicodechar{↔}{\ensuremath{\leftrightarrow}}
\newunicodechar{⦃}{\ensuremath{\{\{}}
\newunicodechar{⦄}{\ensuremath{\}\}}}
\newunicodechar{α}{\ensuremath{\alpha}}
\newunicodechar{β}{\ensuremath{\beta}}
\newunicodechar{ν}{\ensuremath{\nu}}
\newunicodechar{ξ}{\ensuremath{\xi}}
\newunicodechar{ζ}{\ensuremath{\zeta}}
\newunicodechar{φ}{\ensuremath{\phi}}
\usepackage{bbm}
\newunicodechar{𝕒}{\ensuremath{\mathbbm{a}}}
\newunicodechar{𝕓}{\ensuremath{\mathbbm{b}}}
\newunicodechar{𝕔}{\ensuremath{\mathbbm{c}}}
\newunicodechar{𝕕}{\ensuremath{\mathbbm{d}}}
\newunicodechar{𝕩}{\ensuremath{\mathbbm{x}}}
\usepackage{stmaryrd}
\newunicodechar{⦅}{\ensuremath{\llparenthesis}}
\newunicodechar{⦆}{\ensuremath{\rrparenthesis}}
\newunicodechar{≔}{\ensuremath{\coloneqq}}

\title{%
\vspace*{-1cm}%
Nominal techniques as an Agda library\thanks{This work was presented at the \href{https://types2023.webs.upv.es/Accepted.html}{29th International Conference on Types for Proofs and Programs} (TYPES 2023) in Valencia, Spain.  \href{https://types2023.webs.upv.es/slides/S18/TYPES2023-Gabbay-Melkonian.pdf}{Slides}, a \href{https://media.upv.es/\#/portal/video/cbe32610-34aa-11ee-8485-f133f82f8945}{talk recording}, and \href{https://github.com/omelkonian/nominal-agda}{Agda sources on GitHub} are also available.}
\vspace*{-0.1cm}%
}

\author{
Murdoch J. Gabbay\inst{1}
\and
Orestis Melkonian\inst{2}
\vspace*{-0.1cm}%
}

\institute{
  Heriot-Watt University, Scotland
\and
  University of Edinburgh, Scotland
\vspace*{-0.2cm}%
}

\authorrunning{M.J.Gabbay, O.Melkonian}
\titlerunning{Nominal techniques as an Agda library}

\begin{document}

\maketitle
\pagenumbering{gobble}

\paragraph{Introduction}
Nominal techniques~\cite{nominal} provide a mathematically principled approach to dealing
with names and variable binding in programming languages.
However, integrating these ideas in a practical and widespread toolchain has been slow,
and we perceive a chicken-and-egg problem: there are no users for nominal techniques,
because nobody has implemented them, and nobody implements them because there are no users.
This is a pity, but it leaves a positive opportunity to set up a virtuous circle of
broader understanding, adoption, and application of this beautiful technology.

This paper explores an attempt to make nominal techniques accessible
as a library in the Agda proof assistant and programming language~\cite{agda},
which can be viewed as a port of
the first author's Haskell \texttt{nom} package~\cite{nominal-haskell},
although that would be an injustice as its purpose is two-fold:
\begin{enumerate*}
\item provide a convenient library to use nominal techniques in Your Own Agda Formalisation
\item study the meta-theory of nominal techniques in a rigorous and \emph{constructive} way
\end{enumerate*}
A solution to Goal~1 must be ergonomic, meaning that
a \emph{technical} victory of implementing nominal ideas is not enough;
we further require a \emph{moral} victory that the overhead be acceptable for
practical systems.
Apart from this being a literate Agda file,
our results have been mechanised and are publicly accessible:
\url{https://omelkonian.github.io/nominal-agda/}.

\begin{code}[hide]%
\>[0]\AgdaKeyword{open}\AgdaSpace{}%
\AgdaKeyword{import}\AgdaSpace{}%
\AgdaModule{Prelude.Init}\AgdaSymbol{;}\AgdaSpace{}%
\AgdaKeyword{open}\AgdaSpace{}%
\AgdaModule{SetAsType}\<%
\\
\>[0]\AgdaKeyword{open}\AgdaSpace{}%
\AgdaModule{L.Mem}\<%
\\
\>[0]\AgdaKeyword{open}\AgdaSpace{}%
\AgdaKeyword{import}\AgdaSpace{}%
\AgdaModule{Prelude.DecEq}\<%
\\
\>[0]\AgdaKeyword{open}\AgdaSpace{}%
\AgdaKeyword{import}\AgdaSpace{}%
\AgdaModule{Prelude.InfEnumerable}\<%
\\
\>[0]\AgdaKeyword{open}\AgdaSpace{}%
\AgdaKeyword{import}\AgdaSpace{}%
\AgdaModule{Prelude.InferenceRules}\<%
\\
\>[0]\AgdaKeyword{open}\AgdaSpace{}%
\AgdaKeyword{import}\AgdaSpace{}%
\AgdaModule{Prelude.Setoid}\<%
\\
\>[0]\AgdaKeyword{open}\AgdaSpace{}%
\AgdaKeyword{import}\AgdaSpace{}%
\AgdaModule{Prelude.Lists.Dec}\<%
\end{code}

\paragraph{Nominal setup}
We conduct our development under some abstract type of \textbf{atoms},
satisfying certain constraints, namely decidable equality and being infinitely enumerable.\footnote{
  \dots also known as ``unfiniteness'' in a recent nominal mechanization of
  the locally nameless approach~\cite{nominal-locally-nameless}.
}.
We model this in Agda using \emph{module parameters}, which could be instantiated with a
concrete type:

\hspace*{-1cm}
\begin{minipage}{\textwidth}
\begin{code}%
\>[0]\AgdaKeyword{module}\AgdaSpace{}%
\AgdaModule{\AgdaUnderscore{}}\AgdaSpace{}%
\AgdaSymbol{(}\AgdaBound{Atom}\AgdaSpace{}%
\AgdaSymbol{:}\AgdaSpace{}%
\AgdaPrimitive{Type}\AgdaSymbol{)}\AgdaSpace{}%
\AgdaSymbol{⦃}\AgdaSpace{}%
\AgdaBound{\AgdaUnderscore{}}\AgdaSpace{}%
\AgdaSymbol{:}\AgdaSpace{}%
\AgdaRecord{DecEq}\AgdaSpace{}%
\AgdaBound{Atom}\AgdaSpace{}%
\AgdaSymbol{⦄}\AgdaSpace{}%
\AgdaSymbol{⦃}\AgdaSpace{}%
\AgdaBound{\AgdaUnderscore{}}\AgdaSpace{}%
\AgdaSymbol{:}\AgdaSpace{}%
\AgdaRecord{Enumerable∞}\AgdaSpace{}%
\AgdaBound{Atom}\AgdaSpace{}%
\AgdaSymbol{⦄}\AgdaSpace{}%
\AgdaKeyword{where}\<%
\end{code}
\end{minipage}
\begin{code}[hide]%
\>[0]\AgdaKeyword{variable}\AgdaSpace{}%
\AgdaGeneralizable{A}\AgdaSpace{}%
\AgdaSymbol{:}\AgdaSpace{}%
\AgdaPrimitive{Type}\AgdaSymbol{;}\AgdaSpace{}%
\AgdaGeneralizable{𝕒}\AgdaSpace{}%
\AgdaGeneralizable{𝕓}\AgdaSpace{}%
\AgdaGeneralizable{𝕔}\AgdaSpace{}%
\AgdaGeneralizable{𝕕}\AgdaSpace{}%
\AgdaSymbol{:}\AgdaSpace{}%
\AgdaBound{Atom}\AgdaSymbol{;}\AgdaSpace{}%
\AgdaGeneralizable{x}\AgdaSpace{}%
\AgdaSymbol{:}\AgdaSpace{}%
\AgdaGeneralizable{A}\<%
\end{code}
\begin{code}%
\>[0]\AgdaFunction{И}\AgdaSpace{}%
\AgdaSymbol{:}\AgdaSpace{}%
\AgdaSymbol{(}\AgdaBound{Atom}\AgdaSpace{}%
\AgdaSymbol{→}\AgdaSpace{}%
\AgdaPrimitive{Type}\AgdaSymbol{)}\AgdaSpace{}%
\AgdaSymbol{→}\AgdaSpace{}%
\AgdaPrimitive{Type}\<%
\\
\>[0]\AgdaFunction{И}\AgdaSpace{}%
\AgdaBound{φ}\AgdaSpace{}%
\AgdaSymbol{=}\AgdaSpace{}%
\AgdaFunction{∃}\AgdaSpace{}%
\AgdaSymbol{λ}\AgdaSpace{}%
\AgdaSymbol{(}\AgdaBound{xs}\AgdaSpace{}%
\AgdaSymbol{:}\AgdaSpace{}%
\AgdaDatatype{List}\AgdaSpace{}%
\AgdaBound{Atom}\AgdaSymbol{)}\AgdaSpace{}%
\AgdaSymbol{→}\AgdaSpace{}%
\AgdaSymbol{(∀}\AgdaSpace{}%
\AgdaBound{y}\AgdaSpace{}%
\AgdaSymbol{→}\AgdaSpace{}%
\AgdaBound{y}\AgdaSpace{}%
\AgdaOperator{\AgdaFunction{∉}}\AgdaSpace{}%
\AgdaBound{xs}\AgdaSpace{}%
\AgdaSymbol{→}\AgdaSpace{}%
\AgdaBound{φ}\AgdaSpace{}%
\AgdaBound{y}\AgdaSymbol{)}\<%
\end{code}
\begin{code}[hide]%
\>[0]\AgdaFunction{И²}\AgdaSpace{}%
\AgdaSymbol{:}\AgdaSpace{}%
\AgdaSymbol{(}\AgdaBound{Atom}\AgdaSpace{}%
\AgdaSymbol{→}\AgdaSpace{}%
\AgdaBound{Atom}\AgdaSpace{}%
\AgdaSymbol{→}\AgdaSpace{}%
\AgdaPrimitive{Type}\AgdaSymbol{)}\AgdaSpace{}%
\AgdaSymbol{→}\AgdaSpace{}%
\AgdaPrimitive{Type}\<%
\\
\>[0]\AgdaFunction{И²}\AgdaSpace{}%
\AgdaBound{φ}\AgdaSpace{}%
\AgdaSymbol{=}\AgdaSpace{}%
\AgdaFunction{∃}\AgdaSpace{}%
\AgdaSymbol{λ}\AgdaSpace{}%
\AgdaSymbol{(}\AgdaBound{xs}\AgdaSpace{}%
\AgdaSymbol{:}\AgdaSpace{}%
\AgdaDatatype{List}\AgdaSpace{}%
\AgdaBound{Atom}\AgdaSymbol{)}\AgdaSpace{}%
\AgdaSymbol{→}\AgdaSpace{}%
\AgdaSymbol{(∀}\AgdaSpace{}%
\AgdaBound{y}\AgdaSpace{}%
\AgdaBound{z}\AgdaSpace{}%
\AgdaSymbol{→}\AgdaSpace{}%
\AgdaBound{y}\AgdaSpace{}%
\AgdaOperator{\AgdaFunction{∉}}\AgdaSpace{}%
\AgdaBound{xs}\AgdaSpace{}%
\AgdaSymbol{→}\AgdaSpace{}%
\AgdaBound{z}\AgdaSpace{}%
\AgdaOperator{\AgdaFunction{∉}}\AgdaSpace{}%
\AgdaBound{xs}\AgdaSpace{}%
\AgdaSymbol{→}\AgdaSpace{}%
\AgdaBound{φ}\AgdaSpace{}%
\AgdaBound{y}\AgdaSpace{}%
\AgdaBound{z}\AgdaSymbol{)}\<%
\end{code}

\noindent
The И \textbf{quantifier} enforces that a predicate holds for all but finitely many atoms,
and \textbf{swapping} of two atoms can be performed on any type, subject to some laws:

\hspace*{-1.5cm}
\begin{minipage}{.5\textwidth}
\begin{code}%
\>[0]\AgdaKeyword{record}\AgdaSpace{}%
\AgdaRecord{Swap}\AgdaSpace{}%
\AgdaSymbol{(}\AgdaBound{A}\AgdaSpace{}%
\AgdaSymbol{:}\AgdaSpace{}%
\AgdaPrimitive{Type}\AgdaSymbol{)}\AgdaSpace{}%
\AgdaSymbol{:}\AgdaSpace{}%
\AgdaPrimitive{Type}\AgdaSpace{}%
\AgdaKeyword{where}\<%
\\
\>[0][@{}l@{\AgdaIndent{0}}]%
\>[2]\AgdaKeyword{field}\AgdaSpace{}%
\AgdaField{swap}\AgdaSpace{}%
\AgdaSymbol{:}\AgdaSpace{}%
\AgdaBound{Atom}\AgdaSpace{}%
\AgdaSymbol{→}\AgdaSpace{}%
\AgdaBound{Atom}\AgdaSpace{}%
\AgdaSymbol{→}\AgdaSpace{}%
\AgdaBound{A}\AgdaSpace{}%
\AgdaSymbol{→}\AgdaSpace{}%
\AgdaBound{A}\<%
\\
\>[2]\AgdaOperator{\AgdaFunction{⦅\AgdaUnderscore{}↔\AgdaUnderscore{}⦆\AgdaUnderscore{}}}\AgdaSpace{}%
\AgdaSymbol{=}\AgdaSpace{}%
\AgdaField{swap}\<%
\end{code}
\begin{code}[hide]%
\>[0]\AgdaKeyword{open}\AgdaSpace{}%
\AgdaModule{Swap}\AgdaSpace{}%
\AgdaSymbol{⦃...⦄}\<%
\end{code}
\end{minipage}
\vrule
\begin{minipage}{.45\textwidth}
\begin{code}%
\>[0]\AgdaKeyword{instance}\<%
\\
\>[0][@{}l@{\AgdaIndent{0}}]%
\>[2]\AgdaFunction{↔Atom}\AgdaSpace{}%
\AgdaSymbol{:}\AgdaSpace{}%
\AgdaRecord{Swap}\AgdaSpace{}%
\AgdaBound{Atom}\<%
\\
\>[2]\AgdaFunction{↔Atom}\AgdaSpace{}%
\AgdaSymbol{.}\AgdaField{swap}\AgdaSpace{}%
\AgdaBound{x}\AgdaSpace{}%
\AgdaBound{y}\AgdaSpace{}%
\AgdaBound{z}\AgdaSpace{}%
\AgdaSymbol{=}\<%
\\
\>[2][@{}l@{\AgdaIndent{0}}]%
\>[4]\AgdaOperator{\AgdaFunction{if}}\AgdaSpace{}%
\AgdaBound{z}\AgdaSpace{}%
\AgdaOperator{\AgdaFunction{==}}\AgdaSpace{}%
\AgdaBound{x}\AgdaSpace{}%
\AgdaOperator{\AgdaFunction{then}}\AgdaSpace{}%
\AgdaBound{y}\AgdaSpace{}%
\AgdaOperator{\AgdaFunction{else}}\AgdaSpace{}%
\AgdaOperator{\AgdaFunction{if}}\AgdaSpace{}%
\AgdaBound{z}\AgdaSpace{}%
\AgdaOperator{\AgdaFunction{==}}\AgdaSpace{}%
\AgdaBound{y}\AgdaSpace{}%
\AgdaOperator{\AgdaFunction{then}}\AgdaSpace{}%
\AgdaBound{x}\AgdaSpace{}%
\AgdaOperator{\AgdaFunction{else}}\AgdaSpace{}%
\AgdaBound{z}\<%
\end{code}
\begin{code}[hide]%
\>[0]\AgdaKeyword{module}\AgdaSpace{}%
\AgdaModule{\AgdaUnderscore{}}\AgdaSpace{}%
\AgdaSymbol{⦃}\AgdaSpace{}%
\AgdaBound{\AgdaUnderscore{}}\AgdaSpace{}%
\AgdaSymbol{:}\AgdaSpace{}%
\AgdaRecord{Swap}\AgdaSpace{}%
\AgdaGeneralizable{A}\AgdaSpace{}%
\AgdaSymbol{⦄}\AgdaSpace{}%
\AgdaSymbol{\{}\AgdaBound{x}\AgdaSpace{}%
\AgdaSymbol{:}\AgdaSpace{}%
\AgdaGeneralizable{A}\AgdaSymbol{\}}\AgdaSpace{}%
\AgdaKeyword{where}\<%
\end{code}
\end{minipage}
\hspace*{-1.5cm}
\begin{code}%
\>[0][@{}l@{\AgdaIndent{1}}]%
\>[2]\AgdaKeyword{record}\AgdaSpace{}%
\AgdaRecord{SwapLaws}\AgdaSpace{}%
\AgdaSymbol{:}\AgdaSpace{}%
\AgdaPrimitive{Type}\AgdaSpace{}%
\AgdaKeyword{where}\<%
\\
\>[2][@{}l@{\AgdaIndent{0}}]%
\>[4]\AgdaKeyword{field}%
\>[157I]\AgdaField{swap-id}%
\>[21]\AgdaSymbol{:}\AgdaSpace{}%
\AgdaOperator{\AgdaFunction{⦅}}\AgdaSpace{}%
\AgdaGeneralizable{𝕒}\AgdaSpace{}%
\AgdaOperator{\AgdaFunction{↔}}\AgdaSpace{}%
\AgdaGeneralizable{𝕒}\AgdaSpace{}%
\AgdaOperator{\AgdaFunction{⦆}}\AgdaSpace{}%
\AgdaBound{x}\AgdaSpace{}%
\AgdaOperator{\AgdaDatatype{≡}}\AgdaSpace{}%
\AgdaBound{x}\<%
\\
\>[.][@{}l@{}]\<[157I]%
\>[10]\AgdaField{swap-rev}%
\>[21]\AgdaSymbol{:}\AgdaSpace{}%
\AgdaOperator{\AgdaFunction{⦅}}\AgdaSpace{}%
\AgdaGeneralizable{𝕒}\AgdaSpace{}%
\AgdaOperator{\AgdaFunction{↔}}\AgdaSpace{}%
\AgdaGeneralizable{𝕓}\AgdaSpace{}%
\AgdaOperator{\AgdaFunction{⦆}}\AgdaSpace{}%
\AgdaBound{x}\AgdaSpace{}%
\AgdaOperator{\AgdaDatatype{≡}}\AgdaSpace{}%
\AgdaOperator{\AgdaFunction{⦅}}\AgdaSpace{}%
\AgdaGeneralizable{𝕓}\AgdaSpace{}%
\AgdaOperator{\AgdaFunction{↔}}\AgdaSpace{}%
\AgdaGeneralizable{𝕒}\AgdaSpace{}%
\AgdaOperator{\AgdaFunction{⦆}}\AgdaSpace{}%
\AgdaBound{x}\<%
\\
\>[10]\AgdaField{swap-sym}%
\>[21]\AgdaSymbol{:}\AgdaSpace{}%
\AgdaOperator{\AgdaFunction{⦅}}\AgdaSpace{}%
\AgdaGeneralizable{𝕒}\AgdaSpace{}%
\AgdaOperator{\AgdaFunction{↔}}\AgdaSpace{}%
\AgdaGeneralizable{𝕓}\AgdaSpace{}%
\AgdaOperator{\AgdaFunction{⦆}}\AgdaSpace{}%
\AgdaOperator{\AgdaFunction{⦅}}\AgdaSpace{}%
\AgdaGeneralizable{𝕓}\AgdaSpace{}%
\AgdaOperator{\AgdaFunction{↔}}\AgdaSpace{}%
\AgdaGeneralizable{𝕒}\AgdaSpace{}%
\AgdaOperator{\AgdaFunction{⦆}}\AgdaSpace{}%
\AgdaBound{x}\AgdaSpace{}%
\AgdaOperator{\AgdaDatatype{≡}}\AgdaSpace{}%
\AgdaBound{x}\<%
\\
\>[10]\AgdaField{swap-swap}%
\>[21]\AgdaSymbol{:}\AgdaSpace{}%
\AgdaOperator{\AgdaFunction{⦅}}\AgdaSpace{}%
\AgdaGeneralizable{𝕒}\AgdaSpace{}%
\AgdaOperator{\AgdaFunction{↔}}\AgdaSpace{}%
\AgdaGeneralizable{𝕓}\AgdaSpace{}%
\AgdaOperator{\AgdaFunction{⦆}}\AgdaSpace{}%
\AgdaOperator{\AgdaFunction{⦅}}\AgdaSpace{}%
\AgdaGeneralizable{𝕔}\AgdaSpace{}%
\AgdaOperator{\AgdaFunction{↔}}\AgdaSpace{}%
\AgdaGeneralizable{𝕕}\AgdaSpace{}%
\AgdaOperator{\AgdaFunction{⦆}}\AgdaSpace{}%
\AgdaBound{x}\AgdaSpace{}%
\AgdaOperator{\AgdaDatatype{≡}}\AgdaSpace{}%
\AgdaOperator{\AgdaFunction{⦅}}\AgdaSpace{}%
\AgdaOperator{\AgdaFunction{⦅}}\AgdaSpace{}%
\AgdaGeneralizable{𝕒}\AgdaSpace{}%
\AgdaOperator{\AgdaFunction{↔}}\AgdaSpace{}%
\AgdaGeneralizable{𝕓}\AgdaSpace{}%
\AgdaOperator{\AgdaFunction{⦆}}\AgdaSpace{}%
\AgdaGeneralizable{𝕔}\AgdaSpace{}%
\AgdaOperator{\AgdaFunction{↔}}\AgdaSpace{}%
\AgdaOperator{\AgdaFunction{⦅}}\AgdaSpace{}%
\AgdaGeneralizable{𝕒}\AgdaSpace{}%
\AgdaOperator{\AgdaFunction{↔}}\AgdaSpace{}%
\AgdaGeneralizable{𝕓}\AgdaSpace{}%
\AgdaOperator{\AgdaFunction{⦆}}\AgdaSpace{}%
\AgdaGeneralizable{𝕕}\AgdaSpace{}%
\AgdaOperator{\AgdaFunction{⦆}}\AgdaSpace{}%
\AgdaOperator{\AgdaFunction{⦅}}\AgdaSpace{}%
\AgdaGeneralizable{𝕒}\AgdaSpace{}%
\AgdaOperator{\AgdaFunction{↔}}\AgdaSpace{}%
\AgdaGeneralizable{𝕓}\AgdaSpace{}%
\AgdaOperator{\AgdaFunction{⦆}}\AgdaSpace{}%
\AgdaBound{x}\<%
\end{code}

\noindent
We only need to provide instances for the base case of \emph{atoms}
(whence the decidable equality), and \emph{abstractions} (coming up next).
From this we can systematically derive swapping definitions for all user-defined types,
using a compile-time macro/tactic (c.f. the case study later on).

One particularly useful family of axioms in equivariant ZFA foundations~\cite{ezfa}
is that swapping distributes everywhere (constructors, functions, type formers)
with the special case for swapping itself being \AR{swap-swap}.
It is consistent to axiomatize this generalized notion of distributivity for \AR{swap}
and we do so by means of a tactic that realises this \emph{axiom scheme}.
Most of the time we are working with types that have \textbf{finite support},
expressed using the 'new' quantifier:
\begin{code}[hide]%
\>[2]\AgdaFunction{FinSupp}\AgdaSpace{}%
\AgdaSymbol{:}\AgdaSpace{}%
\AgdaBound{A}\AgdaSpace{}%
\AgdaSymbol{→}\AgdaSpace{}%
\AgdaPrimitive{Type}\<%
\\
\>[2]\AgdaFunction{FinSupp}\AgdaSpace{}%
\AgdaBound{x}\AgdaSpace{}%
\AgdaSymbol{=}\<%
\end{code}
\begin{code}[inline]%
\>[2][@{}l@{\AgdaIndent{1}}]%
\>[4]\AgdaFunction{И²}\AgdaSpace{}%
\AgdaSymbol{λ}\AgdaSpace{}%
\AgdaBound{𝕒}\AgdaSpace{}%
\AgdaBound{𝕓}\AgdaSpace{}%
\AgdaSymbol{→}\AgdaSpace{}%
\AgdaField{swap}\AgdaSpace{}%
\AgdaBound{𝕓}\AgdaSpace{}%
\AgdaBound{𝕒}\AgdaSpace{}%
\AgdaBound{x}\AgdaSpace{}%
\AgdaOperator{\AgdaDatatype{≡}}\AgdaSpace{}%
\AgdaBound{x}\<%
\end{code}
\begin{code}[hide]%
\>[2]\AgdaFunction{Equivariant}\AgdaSpace{}%
\AgdaSymbol{:}\AgdaSpace{}%
\AgdaBound{A}\AgdaSpace{}%
\AgdaSymbol{→}\AgdaSpace{}%
\AgdaPrimitive{Type}\<%
\\
\>[2]\AgdaFunction{Equivariant}\AgdaSpace{}%
\AgdaBound{x}\AgdaSpace{}%
\AgdaSymbol{=}\AgdaSpace{}%
\AgdaFunction{∃}\AgdaSpace{}%
\AgdaSymbol{λ}\AgdaSpace{}%
\AgdaSymbol{(}\AgdaBound{p}\AgdaSpace{}%
\AgdaSymbol{:}\AgdaSpace{}%
\AgdaFunction{FinSupp}\AgdaSpace{}%
\AgdaBound{x}\AgdaSymbol{)}\AgdaSpace{}%
\AgdaSymbol{→}\AgdaSpace{}%
\AgdaBound{p}\AgdaSpace{}%
\AgdaSymbol{.}\AgdaField{proj₁}\AgdaSpace{}%
\AgdaOperator{\AgdaDatatype{≡}}\AgdaSpace{}%
\AgdaInductiveConstructor{[]}\<%
\end{code}
. We can then define \textbf{equivariant} elements that admit the empty support,
as well as an operation to generate fresh atoms
\begin{code}[hide]%
\>[2]\AgdaKeyword{postulate}\<%
\end{code}
\begin{code}[inline]%
\>[2][@{}l@{\AgdaIndent{1}}]%
\>[4]\AgdaPostulate{freshAtom}\AgdaSpace{}%
\AgdaSymbol{:}\AgdaSpace{}%
\AgdaBound{A}\AgdaSpace{}%
\AgdaSymbol{→}\AgdaSpace{}%
\AgdaBound{Atom}\<%
\end{code}
~(whence the module requirement that atoms are infinitely enumerable).
Agda is constructive, so \AF{freshAtom} is constructive too,
which is different from how fresh atoms are used in (non-constructive) set theories.
An \textbf{abstraction} is just a pair of an atom and an element:

\vspace{1pt}
\hspace*{-1cm}
\begin{minipage}{.4\textwidth}
\begin{code}[hide]%
\>[2]\AgdaFunction{Abs}\AgdaSpace{}%
\AgdaSymbol{:}\AgdaSpace{}%
\AgdaPrimitive{Type}\AgdaSpace{}%
\AgdaSymbol{→}\AgdaSpace{}%
\AgdaPrimitive{Type}\<%
\end{code}
\begin{code}%
\>[2]\AgdaFunction{Abs}\AgdaSpace{}%
\AgdaBound{A}\AgdaSpace{}%
\AgdaSymbol{=}\AgdaSpace{}%
\AgdaBound{Atom}\AgdaSpace{}%
\AgdaOperator{\AgdaFunction{×}}\AgdaSpace{}%
\AgdaBound{A}\<%
\\
\\[\AgdaEmptyExtraSkip]%
\>[2]\AgdaFunction{conc}\AgdaSpace{}%
\AgdaSymbol{:}\AgdaSpace{}%
\AgdaFunction{Abs}\AgdaSpace{}%
\AgdaBound{A}\AgdaSpace{}%
\AgdaSymbol{→}\AgdaSpace{}%
\AgdaBound{Atom}\AgdaSpace{}%
\AgdaSymbol{→}\AgdaSpace{}%
\AgdaBound{A}\<%
\\
\>[2]\AgdaFunction{conc}\AgdaSpace{}%
\AgdaSymbol{(}\AgdaBound{𝕒}\AgdaSpace{}%
\AgdaOperator{\AgdaInductiveConstructor{,}}\AgdaSpace{}%
\AgdaBound{x}\AgdaSymbol{)}\AgdaSpace{}%
\AgdaBound{𝕓}\AgdaSpace{}%
\AgdaSymbol{=}\AgdaSpace{}%
\AgdaField{swap}\AgdaSpace{}%
\AgdaBound{𝕓}\AgdaSpace{}%
\AgdaBound{𝕒}\AgdaSpace{}%
\AgdaBound{x}\<%
\end{code}
\end{minipage}
\vrule
\hspace*{-.5cm}
\begin{minipage}{.55\textwidth}
\begin{code}%
\>[2]\AgdaKeyword{instance}\<%
\\
\>[2][@{}l@{\AgdaIndent{0}}]%
\>[4]\AgdaFunction{↔Abs}\AgdaSpace{}%
\AgdaSymbol{:}\AgdaSpace{}%
\AgdaRecord{Swap}\AgdaSpace{}%
\AgdaSymbol{(}\AgdaFunction{Abs}\AgdaSpace{}%
\AgdaBound{A}\AgdaSymbol{)}\<%
\\
\>[4]\AgdaFunction{↔Abs}\AgdaSpace{}%
\AgdaSymbol{.}\AgdaField{swap}\AgdaSpace{}%
\AgdaBound{𝕒}\AgdaSpace{}%
\AgdaBound{𝕓}\AgdaSpace{}%
\AgdaSymbol{(}\AgdaBound{𝕔}\AgdaSpace{}%
\AgdaOperator{\AgdaInductiveConstructor{,}}\AgdaSpace{}%
\AgdaBound{x}\AgdaSymbol{)}\AgdaSpace{}%
\AgdaSymbol{=}\AgdaSpace{}%
\AgdaSymbol{(}\AgdaField{swap}\AgdaSpace{}%
\AgdaBound{𝕒}\AgdaSpace{}%
\AgdaBound{𝕓}\AgdaSpace{}%
\AgdaBound{𝕔}\AgdaSpace{}%
\AgdaOperator{\AgdaInductiveConstructor{,}}\AgdaSpace{}%
\AgdaField{swap}\AgdaSpace{}%
\AgdaBound{𝕒}\AgdaSpace{}%
\AgdaBound{𝕓}\AgdaSpace{}%
\AgdaBound{x}\AgdaSymbol{)}\<%
\end{code}
\end{minipage}
\vspace{1pt}

\noindent
Note that we can also provide a \emph{correct-by-construction} and \emph{total} concretion function.
In nominal techniques based on Fraenkel-Mostowski set theory~\cite{nominal} this is impossible,
and it seems to be a novel observation that in a constructive setup a total concretion function is fine.

\begin{code}[hide]%
\>[0]\AgdaKeyword{open}\AgdaSpace{}%
\AgdaKeyword{import}\AgdaSpace{}%
\AgdaModule{Prelude.Init}\AgdaSpace{}%
\AgdaKeyword{hiding}\AgdaSpace{}%
\AgdaSymbol{(}\AgdaOperator{\AgdaInductiveConstructor{[\AgdaUnderscore{}]}}\AgdaSymbol{)}\AgdaSpace{}%
\AgdaKeyword{renaming}\AgdaSpace{}%
\AgdaSymbol{(}\AgdaOperator{\AgdaFunction{\AgdaUnderscore{}++\AgdaUnderscore{}}}\AgdaSpace{}%
\AgdaSymbol{to}\AgdaSpace{}%
\AgdaOperator{\AgdaFunction{\AgdaUnderscore{}◇\AgdaUnderscore{}}}\AgdaSymbol{);}\AgdaSpace{}%
\AgdaKeyword{open}\AgdaSpace{}%
\AgdaModule{SetAsType}\<%
\\
\>[0]\AgdaKeyword{open}\AgdaSpace{}%
\AgdaModule{L.Mem}\<%
\\
\>[0]\AgdaKeyword{open}\AgdaSpace{}%
\AgdaKeyword{import}\AgdaSpace{}%
\AgdaModule{Prelude.DecEq}\<%
\\
\>[0]\AgdaKeyword{open}\AgdaSpace{}%
\AgdaKeyword{import}\AgdaSpace{}%
\AgdaModule{Prelude.InfEnumerable}\<%
\\
\>[0]\AgdaKeyword{open}\AgdaSpace{}%
\AgdaKeyword{import}\AgdaSpace{}%
\AgdaModule{Prelude.InferenceRules}\<%
\\
\>[0]\AgdaKeyword{open}\AgdaSpace{}%
\AgdaKeyword{import}\AgdaSpace{}%
\AgdaModule{Prelude.Lists.Dec}\<%
\\
\\[\AgdaEmptyExtraSkip]%
\>[0]\AgdaKeyword{module}\AgdaSpace{}%
\AgdaModule{\AgdaUnderscore{}}\AgdaSpace{}%
\AgdaSymbol{(}\AgdaBound{Atom}\AgdaSpace{}%
\AgdaSymbol{:}\AgdaSpace{}%
\AgdaPrimitive{Type}\AgdaSymbol{)}\AgdaSpace{}%
\AgdaSymbol{⦃}\AgdaSpace{}%
\AgdaBound{\AgdaUnderscore{}}\AgdaSpace{}%
\AgdaSymbol{:}\AgdaSpace{}%
\AgdaRecord{DecEq}\AgdaSpace{}%
\AgdaBound{Atom}\AgdaSpace{}%
\AgdaSymbol{⦄}\AgdaSpace{}%
\AgdaSymbol{⦃}\AgdaSpace{}%
\AgdaBound{\AgdaUnderscore{}}\AgdaSpace{}%
\AgdaSymbol{:}\AgdaSpace{}%
\AgdaRecord{Enumerable∞}\AgdaSpace{}%
\AgdaBound{Atom}\AgdaSpace{}%
\AgdaSymbol{⦄}\AgdaSpace{}%
\AgdaKeyword{where}\<%
\end{code}

\paragraph{Case study}

\begin{code}[hide]%
\>[0]\AgdaKeyword{open}\AgdaSpace{}%
\AgdaKeyword{import}\AgdaSpace{}%
\AgdaModule{Nominal}\AgdaSpace{}%
\AgdaBound{Atom}\<%
\end{code}
Once equipped with all expected nominal facilities,
in particular \emph{atoms} and \emph{atom abstractions},
it is easy to define terms in \textbf{untyped λ-calculus} without mentioning de Bruijn indices
or anything of that sort.
For the sake of ergonomics and efficient theorem proving, we provide a meta-programming
macro --- based on \emph{elaborator reflection}~\cite{elab} --- that is able to automatically
derive the implementation of swapping of any type based on its structure.

\vspace{1pt}
\hspace*{-1.5cm}
\begin{minipage}{.5\textwidth}
\begin{code}%
\>[0]\AgdaKeyword{data}\AgdaSpace{}%
\AgdaDatatype{Term}\AgdaSpace{}%
\AgdaSymbol{:}\AgdaSpace{}%
\AgdaPrimitive{Type}\AgdaSpace{}%
\AgdaKeyword{where}\<%
\\
\>[0][@{}l@{\AgdaIndent{0}}]%
\>[2]\AgdaOperator{\AgdaInductiveConstructor{`\AgdaUnderscore{}}}%
\>[7]\AgdaSymbol{:}\AgdaSpace{}%
\AgdaBound{Atom}\AgdaSpace{}%
\AgdaSymbol{→}\AgdaSpace{}%
\AgdaDatatype{Term}\<%
\\
\>[2]\AgdaOperator{\AgdaInductiveConstructor{\AgdaUnderscore{}·\AgdaUnderscore{}}}%
\>[7]\AgdaSymbol{:}\AgdaSpace{}%
\AgdaDatatype{Term}\AgdaSpace{}%
\AgdaSymbol{→}\AgdaSpace{}%
\AgdaDatatype{Term}\AgdaSpace{}%
\AgdaSymbol{→}\AgdaSpace{}%
\AgdaDatatype{Term}\<%
\\
\>[2]\AgdaOperator{\AgdaInductiveConstructor{ƛ\AgdaUnderscore{}}}%
\>[7]\AgdaSymbol{:}\AgdaSpace{}%
\AgdaRecord{Abs}\AgdaSpace{}%
\AgdaDatatype{Term}\AgdaSpace{}%
\AgdaSymbol{→}\AgdaSpace{}%
\AgdaDatatype{Term}\<%
\end{code}
\begin{code}[hide]%
\>[0]\AgdaSymbol{\{-\#}\AgdaSpace{}%
\AgdaKeyword{TERMINATING}\AgdaSpace{}%
\AgdaSymbol{\#-\}}\<%
\end{code}
\begin{code}%
\>[0]\AgdaKeyword{unquoteDecl}\AgdaSpace{}%
\AgdaFunction{↔Term}\AgdaSpace{}%
\AgdaSymbol{=}\<%
\end{code}
\begin{code}[hide,inline]%
\>[0][@{}l@{\AgdaIndent{1}}]%
\>[2]\AgdaKeyword{let}\AgdaSpace{}%
\AgdaKeyword{open}\AgdaSpace{}%
\AgdaModule{L}\AgdaSpace{}%
\AgdaKeyword{in}\<%
\end{code}
\hspace*{1cm}
\begin{code}[inline]%
\>[2]\AgdaMacro{DERIVE}\AgdaSpace{}%
\AgdaRecord{Swap}\AgdaSpace{}%
\AgdaOperator{\AgdaFunction{[}}\AgdaSpace{}%
\AgdaKeyword{quote}\AgdaSpace{}%
\AgdaDatatype{Term}\AgdaSpace{}%
\AgdaOperator{\AgdaInductiveConstructor{,}}\AgdaSpace{}%
\AgdaFunction{↔Term}\AgdaSpace{}%
\AgdaOperator{\AgdaFunction{]}}\<%
\end{code}
\begin{code}[hide]%
\>[0]\AgdaKeyword{infix}%
\>[7]\AgdaNumber{30}\AgdaSpace{}%
\AgdaOperator{\AgdaInductiveConstructor{`\AgdaUnderscore{}}}\<%
\\
\>[0]\AgdaKeyword{infixl}\AgdaSpace{}%
\AgdaNumber{20}\AgdaSpace{}%
\AgdaOperator{\AgdaInductiveConstructor{\AgdaUnderscore{}·\AgdaUnderscore{}}}\<%
\\
\>[0]\AgdaKeyword{infixr}\AgdaSpace{}%
\AgdaNumber{10}\AgdaSpace{}%
\AgdaOperator{\AgdaInductiveConstructor{ƛ\AgdaUnderscore{}}}\<%
\\
\>[0]\AgdaKeyword{infixr}\AgdaSpace{}%
\AgdaNumber{5}\AgdaSpace{}%
\AgdaOperator{\AgdaInductiveConstructor{ƛ\AgdaUnderscore{}⇒\AgdaUnderscore{}}}\<%
\\
\>[0]\AgdaKeyword{pattern}\AgdaSpace{}%
\AgdaOperator{\AgdaInductiveConstructor{ƛ\AgdaUnderscore{}⇒\AgdaUnderscore{}}}\AgdaSpace{}%
\AgdaBound{x}\AgdaSpace{}%
\AgdaBound{y}\AgdaSpace{}%
\AgdaSymbol{=}\AgdaSpace{}%
\AgdaOperator{\AgdaInductiveConstructor{ƛ}}\AgdaSpace{}%
\AgdaInductiveConstructor{abs}\AgdaSpace{}%
\AgdaBound{x}\AgdaSpace{}%
\AgdaBound{y}\<%
\\
\>[0]\AgdaKeyword{variable}\<%
\\
\>[0][@{}l@{\AgdaIndent{0}}]%
\>[2]\AgdaGeneralizable{x}\AgdaSpace{}%
\AgdaGeneralizable{y}\AgdaSpace{}%
\AgdaGeneralizable{𝕒}\AgdaSpace{}%
\AgdaGeneralizable{𝕓}\AgdaSpace{}%
\AgdaGeneralizable{𝕔}\AgdaSpace{}%
\AgdaGeneralizable{𝕕}\AgdaSpace{}%
\AgdaSymbol{:}\AgdaSpace{}%
\AgdaBound{Atom}\<%
\\
\>[2]\AgdaGeneralizable{t}\AgdaSpace{}%
\AgdaGeneralizable{t'}\AgdaSpace{}%
\AgdaGeneralizable{t″}\AgdaSpace{}%
\AgdaGeneralizable{L}\AgdaSpace{}%
\AgdaGeneralizable{L'}\AgdaSpace{}%
\AgdaGeneralizable{M}\AgdaSpace{}%
\AgdaGeneralizable{M'}\AgdaSpace{}%
\AgdaGeneralizable{N}\AgdaSpace{}%
\AgdaGeneralizable{N'}\AgdaSpace{}%
\AgdaGeneralizable{M₁}\AgdaSpace{}%
\AgdaGeneralizable{M₂}\AgdaSpace{}%
\AgdaSymbol{:}\AgdaSpace{}%
\AgdaDatatype{Term}\<%
\\
\>[2]\AgdaGeneralizable{f}\AgdaSpace{}%
\AgdaGeneralizable{g}\AgdaSpace{}%
\AgdaSymbol{:}\AgdaSpace{}%
\AgdaRecord{Abs}\AgdaSpace{}%
\AgdaDatatype{Term}\<%
\\
\>[0]\AgdaKeyword{infix}\AgdaSpace{}%
\AgdaNumber{0}\AgdaSpace{}%
\AgdaOperator{\AgdaDatatype{\AgdaUnderscore{}≈\AgdaUnderscore{}}}\<%
\end{code}
\end{minipage}
\hspace{-.3cm}
\vrule
\hspace{-.7cm}
\begin{minipage}{.45\textwidth}
\begin{code}%
\>[0]\AgdaKeyword{data}\AgdaSpace{}%
\AgdaOperator{\AgdaDatatype{\AgdaUnderscore{}≈\AgdaUnderscore{}}}\AgdaSpace{}%
\AgdaSymbol{:}\AgdaSpace{}%
\AgdaDatatype{Term}\AgdaSpace{}%
\AgdaSymbol{→}\AgdaSpace{}%
\AgdaDatatype{Term}\AgdaSpace{}%
\AgdaSymbol{→}\AgdaSpace{}%
\AgdaPrimitive{Type}\AgdaSpace{}%
\AgdaKeyword{where}\<%
\\
\>[0][@{}l@{\AgdaIndent{0}}]%
\>[2]\AgdaInductiveConstructor{ν≈}\AgdaSpace{}%
\AgdaSymbol{:}\AgdaSpace{}%
\AgdaOperator{\AgdaInductiveConstructor{`}}\AgdaSpace{}%
\AgdaGeneralizable{x}\AgdaSpace{}%
\AgdaOperator{\AgdaDatatype{≈}}\AgdaSpace{}%
\AgdaOperator{\AgdaInductiveConstructor{`}}\AgdaSpace{}%
\AgdaGeneralizable{x}\<%
\\
\>[2]\AgdaInductiveConstructor{ξ≈}\AgdaSpace{}%
\AgdaSymbol{:}\AgdaSpace{}%
\AgdaGeneralizable{L}\AgdaSpace{}%
\AgdaOperator{\AgdaDatatype{≈}}\AgdaSpace{}%
\AgdaGeneralizable{L'}\AgdaSpace{}%
\AgdaSymbol{→}\AgdaSpace{}%
\AgdaGeneralizable{M}\AgdaSpace{}%
\AgdaOperator{\AgdaDatatype{≈}}\AgdaSpace{}%
\AgdaGeneralizable{M'}\AgdaSpace{}%
\AgdaSymbol{→}\AgdaSpace{}%
\AgdaGeneralizable{L}\AgdaSpace{}%
\AgdaOperator{\AgdaInductiveConstructor{·}}\AgdaSpace{}%
\AgdaGeneralizable{M}\AgdaSpace{}%
\AgdaOperator{\AgdaDatatype{≈}}\AgdaSpace{}%
\AgdaGeneralizable{L'}\AgdaSpace{}%
\AgdaOperator{\AgdaInductiveConstructor{·}}\AgdaSpace{}%
\AgdaGeneralizable{M'}\<%
\\
\>[2]\AgdaInductiveConstructor{ζ≈}\AgdaSpace{}%
\AgdaSymbol{:}\AgdaSpace{}%
\AgdaFunction{И}\AgdaSpace{}%
\AgdaSymbol{(λ}\AgdaSpace{}%
\AgdaBound{𝕩}\AgdaSpace{}%
\AgdaSymbol{→}\AgdaSpace{}%
\AgdaFunction{conc}\AgdaSpace{}%
\AgdaGeneralizable{f}\AgdaSpace{}%
\AgdaBound{𝕩}\AgdaSpace{}%
\AgdaOperator{\AgdaDatatype{≈}}\AgdaSpace{}%
\AgdaFunction{conc}\AgdaSpace{}%
\AgdaGeneralizable{g}\AgdaSpace{}%
\AgdaBound{𝕩}\AgdaSymbol{)}\AgdaSpace{}%
\AgdaSymbol{→}\AgdaSpace{}%
\AgdaOperator{\AgdaInductiveConstructor{ƛ}}\AgdaSpace{}%
\AgdaGeneralizable{f}\AgdaSpace{}%
\AgdaOperator{\AgdaDatatype{≈}}\AgdaSpace{}%
\AgdaOperator{\AgdaInductiveConstructor{ƛ}}\AgdaSpace{}%
\AgdaGeneralizable{g}\<%
\end{code}
\begin{code}[hide]%
\>[0]\AgdaKeyword{open}\AgdaSpace{}%
\AgdaKeyword{import}\AgdaSpace{}%
\AgdaModule{Prelude.Setoid}\<%
\\
\>[0]\AgdaKeyword{instance}\AgdaSpace{}%
\AgdaKeyword{postulate}\<%
\\
\>[0][@{}l@{\AgdaIndent{0}}]%
\>[2]\AgdaPostulate{\AgdaUnderscore{}}\AgdaSpace{}%
\AgdaSymbol{:}\AgdaSpace{}%
\AgdaRecord{ISetoid}\AgdaSpace{}%
\AgdaDatatype{Term}\<%
\\
\>[2]\AgdaPostulate{\AgdaUnderscore{}}\AgdaSpace{}%
\AgdaSymbol{:}\AgdaSpace{}%
\AgdaRecord{SetoidLaws}\AgdaSpace{}%
\AgdaDatatype{Term}\<%
\\
\>[2]\AgdaPostulate{\AgdaUnderscore{}}\AgdaSpace{}%
\AgdaSymbol{:}\AgdaSpace{}%
\AgdaRecord{SwapLaws}\AgdaSpace{}%
\AgdaDatatype{Term}\<%
\\
\>[2]\AgdaPostulate{\AgdaUnderscore{}}\AgdaSpace{}%
\AgdaSymbol{:}\AgdaSpace{}%
\AgdaRecord{FinitelySupported}\AgdaSpace{}%
\AgdaDatatype{Term}\<%
\end{code}
\end{minipage}
\vspace{1pt}

\noindent
We can naturally express α-equivalence of λ-terms using the И quantifier
and manually prove the aforementioned swapping laws and
the fact that every λ-term has finite support.
However, these all admit a systematic datatype-generic construction and
we are currently in the process of automating them.
The rest of the development remains identical to the mechanization presented in the
PLFA textbook~\cite{plfa}, particularly the `Untyped' chapter.
Meanwhile, the gnarly `Substitution' appendix involving tedious index manipulations
is now replaced by the usual nominal presentation of substitution,
alongside a few general lemmas about equivariance and support:

\vspace{1pt}
\hspace*{-1.5ex}\begin{code}[hide]%
\>[0]\AgdaKeyword{infix}\AgdaSpace{}%
\AgdaNumber{30}\AgdaSpace{}%
\AgdaOperator{\AgdaFunction{\AgdaUnderscore{}[\AgdaUnderscore{}≔\AgdaUnderscore{}]}}\<%
\\
\>[0]\AgdaSymbol{\{-\#}\AgdaSpace{}%
\AgdaKeyword{TERMINATING}\AgdaSpace{}%
\AgdaSymbol{\#-\}}\<%
\end{code}
\hspace*{-1cm}
\begin{minipage}{\textwidth}
\begin{code}%
\>[0]\AgdaOperator{\AgdaFunction{\AgdaUnderscore{}[\AgdaUnderscore{}≔\AgdaUnderscore{}]}}\AgdaSpace{}%
\AgdaSymbol{:}\AgdaSpace{}%
\AgdaDatatype{Term}\AgdaSpace{}%
\AgdaSymbol{→}\AgdaSpace{}%
\AgdaBound{Atom}\AgdaSpace{}%
\AgdaSymbol{→}\AgdaSpace{}%
\AgdaDatatype{Term}\AgdaSpace{}%
\AgdaSymbol{→}\AgdaSpace{}%
\AgdaDatatype{Term}\<%
\\
\>[0]\AgdaSymbol{(}\AgdaOperator{\AgdaInductiveConstructor{`}}\AgdaSpace{}%
\AgdaBound{x}\AgdaSymbol{)}%
\>[9]\AgdaOperator{\AgdaFunction{[}}\AgdaSpace{}%
\AgdaBound{𝕒}\AgdaSpace{}%
\AgdaOperator{\AgdaFunction{≔}}\AgdaSpace{}%
\AgdaBound{N}\AgdaSpace{}%
\AgdaOperator{\AgdaFunction{]}}%
\>[20]\AgdaSymbol{=}%
\>[23]\AgdaOperator{\AgdaFunction{if}}\AgdaSpace{}%
\AgdaBound{x}\AgdaSpace{}%
\AgdaOperator{\AgdaFunction{==}}\AgdaSpace{}%
\AgdaBound{𝕒}\AgdaSpace{}%
\AgdaOperator{\AgdaFunction{then}}\AgdaSpace{}%
\AgdaBound{N}\AgdaSpace{}%
\AgdaOperator{\AgdaFunction{else}}\AgdaSpace{}%
\AgdaOperator{\AgdaInductiveConstructor{`}}\AgdaSpace{}%
\AgdaBound{x}\<%
\\
\>[0]\AgdaSymbol{(}\AgdaBound{L}\AgdaSpace{}%
\AgdaOperator{\AgdaInductiveConstructor{·}}\AgdaSpace{}%
\AgdaBound{M}\AgdaSymbol{)}%
\>[9]\AgdaOperator{\AgdaFunction{[}}\AgdaSpace{}%
\AgdaBound{𝕒}\AgdaSpace{}%
\AgdaOperator{\AgdaFunction{≔}}\AgdaSpace{}%
\AgdaBound{N}\AgdaSpace{}%
\AgdaOperator{\AgdaFunction{]}}%
\>[20]\AgdaSymbol{=}%
\>[23]\AgdaBound{L}\AgdaSpace{}%
\AgdaOperator{\AgdaFunction{[}}\AgdaSpace{}%
\AgdaBound{𝕒}\AgdaSpace{}%
\AgdaOperator{\AgdaFunction{≔}}\AgdaSpace{}%
\AgdaBound{N}\AgdaSpace{}%
\AgdaOperator{\AgdaFunction{]}}\AgdaSpace{}%
\AgdaOperator{\AgdaInductiveConstructor{·}}\AgdaSpace{}%
\AgdaBound{M}\AgdaSpace{}%
\AgdaOperator{\AgdaFunction{[}}\AgdaSpace{}%
\AgdaBound{𝕒}\AgdaSpace{}%
\AgdaOperator{\AgdaFunction{≔}}\AgdaSpace{}%
\AgdaBound{N}\AgdaSpace{}%
\AgdaOperator{\AgdaFunction{]}}\<%
\\
\>[0]\AgdaSymbol{(}\AgdaOperator{\AgdaInductiveConstructor{ƛ}}\AgdaSpace{}%
\AgdaBound{f}\AgdaSymbol{)}%
\>[9]\AgdaOperator{\AgdaFunction{[}}\AgdaSpace{}%
\AgdaBound{𝕒}\AgdaSpace{}%
\AgdaOperator{\AgdaFunction{≔}}\AgdaSpace{}%
\AgdaBound{N}\AgdaSpace{}%
\AgdaOperator{\AgdaFunction{]}}%
\>[20]\AgdaSymbol{=}%
\>[23]\AgdaOperator{\AgdaInductiveConstructor{ƛ}}\AgdaSpace{}%
\AgdaFunction{z}\AgdaSpace{}%
\AgdaOperator{\AgdaInductiveConstructor{⇒}}\AgdaSpace{}%
\AgdaFunction{conc}\AgdaSpace{}%
\AgdaBound{f}\AgdaSpace{}%
\AgdaFunction{z}\AgdaSpace{}%
\AgdaOperator{\AgdaFunction{[}}\AgdaSpace{}%
\AgdaBound{𝕒}\AgdaSpace{}%
\AgdaOperator{\AgdaFunction{≔}}\AgdaSpace{}%
\AgdaBound{N}\AgdaSpace{}%
\AgdaOperator{\AgdaFunction{]}}\AgdaSpace{}%
\AgdaKeyword{where}\AgdaSpace{}%
\AgdaFunction{z}\AgdaSpace{}%
\AgdaSymbol{=}\AgdaSpace{}%
\AgdaFunction{freshAtom}\AgdaSpace{}%
\AgdaSymbol{(}\AgdaBound{𝕒}\AgdaSpace{}%
\AgdaOperator{\AgdaInductiveConstructor{∷}}\AgdaSpace{}%
\AgdaFunction{supp}\AgdaSpace{}%
\AgdaBound{f}\AgdaSpace{}%
\AgdaOperator{\AgdaFunction{◇}}\AgdaSpace{}%
\AgdaFunction{supp}\AgdaSpace{}%
\AgdaBound{N}\AgdaSymbol{)}\<%
\end{code}
\end{minipage}

\noindent
We still have a few remaining lemmas to prove to fully cover the PLFA chapter
on untyped λ-calculus,
but we do not see any inherent obstacles to completing the confluence proof.
A good next step would be to formalise a proof of \emph{cut elimination} for
first-order logic, since this involves name-abstraction on both terms and proof-trees.


\paragraph{Related work}
There have been previous nominal mechanizations in Agda that focus
on the concrete instance of the untyped λ-calculus and include a proof of confluence
~\cite{alpha-agda,alpha-agda-confluence}.
Ours closely matches the non-mechanized formulation in~\cite{nominal-lambda},
which the Haskell \texttt{nom} package~\cite{nominal-haskell} then implements.
Another representation of nominal sets in Agda~\cite{nominal-agda-msc} is preliminary
and we would hope that our approach is more ergonomic and more amenable to scaling up.
We treat our Agda library as a complement to other nominal implementations
(in FreshML~\cite{nominal-ml}, Isabelle/HOL~\cite{nominal-isabelle}, and Nuprl~\cite{nominal-nuprl})
that is ergonomic, lightweight, accessible, and illustrates
the practical compatibility of nominal techniques within a constructive type system.

\vspace{-.35cm}
\bibliographystyle{plainurl}
\bibliography{sources}
\end{document}